# Secure Biometric-based Remote Authentication Protocol using Chebyshev Polynomials and Fuzzy Extractor

Thi Ai Thao Nguyen
*Ho Chi Minh University of Technology, VNU-HCM*
Ho Chi Minh, Vietnam
thaonguyen@hcmut.edu.vn

Tran Khanh Dang
*Ho Chi Minh University of Technology, VNU-HCM*
Ho Chi Minh, Vietnam
khanh@hcmut.edu.vn

Quynh Chi Truong
*Ho Chi Minh University of Technology, VNU-HCM*
Ho Chi Minh, Vietnam
tqchi@hcmut.edu.vn

Dinh Thanh Nguyen
*Ho Chi Minh University of Technology, VNU-HCM*
Ho Chi Minh, Vietnam
dinhthanh@hcmut.edu.vn

*Abstract*— Nowadays, biometric-based authentication method has been gradually replacing the traditional authentication username/password. Using biometric data for authentication is very convenience, however some issues related to security and privacy of this kind of system need to be taken into consideration carefully. In this paper, we have proposed a multi factor biometric-based remote authentication protocol. This protocol not only is resistant against the attacks over the insecure network, but also protects user's sensitive data such as biometric template stored in the database, thanks to the combination of Chebyshev polynomials and fuzzy extractor. Our proposal overcomes the vulnerabilities of some previous works. At the same time, the protocol also obtains a low false accept rate (FAR) and false reject rate (FRR) of 3.07% and 7.69%, respectively.

## I. INTRODUCTION

In the 21st century, we have been witnessing the tremendous growth of information technology in both quantity and quality. Research results from the modern technology are increasingly applied in people.'s daily life. Among them, Internet is one of the greatest accomplishments that connects people all over the world. It has been widely used in every aspect of life. The widespread use of the Internet has been the foundation of the global trade and also the applications. of electronic commerce. The authentication of each individual over the network imposes many risks. Therefore, the demand for a secure information system, which ensures the confidentiality of the communication between client and server becomes more and more important. The first step to build a secure information system is authentication.

Traditional authentication method that most e-commerce providers are using is username/password. To login system, users have to provide their username and corresponding password. However, the password itself has some natural setbacks. Firstly, it cannot identify legal user with an imposter who is able to access to user's password. Besides, the more complicated – more secured a password is, the harder it is for users to remember. That is to say, a "true" password is difficult for people to remember but easy for computer to figure out. Especially, with recent technology development, computer ability is being enhanced; meaning password cracking chance is rising too. For that reason, biometric based authentication method was born. The first advantage to be mentioned is that biometric (such as face, voice, iris, fingerprint, palm-print, gait, signature,…) reflects a specific individual which helps preventing multi-user usage from one account [1]. Moreover, using biometric method is more convenient for users since they do not have to remember or carry it with them.

However, biometrics still has some troubles with the security of the authentication system. As we all know, human has a limited number of biometric traits; therefore, we cannot change our biometrics frequently like password once we suspect that the templates are revealed [2]. Moreover, the fact that people register many online services makes them use the same biometrics to sign in some services. That leads to the cross-matching attacks when attackers follow user's biometric template cross the online services in order to track their activities. Another concern relates to the natural set-backs of biometrics. The fact that biometrics reflects a specific individual means it contains sensitive information which users do not want attacker or even the server storing users' authentication data to discover. Last but not least, the network security needs to be discussed when user's private information is transmitted over insecure network [3].

The goal of this work is to propose the biometric-based remote authentication protocol relied on Chebyshev polynomials and Fuzzy Extractor. This protocol not only protects sensitive data of users from attackers inside and outside system but also gain the facial recognition rate high. That makes the proposal highly practical.

The remaining parts of this paper are organized as follows. In the section 2, related works is briefly reviewed. We show what previous works have done and their limitations. From that point, we present our motivation to fill the gap. In section 3, we introduce the preliminaries used in the proposal. In the next section, our proposed protocol is described in detail. In the section 5, the evaluation is presented to demonstrate for our proposal. Finally, the conclusion is included in the section 6.

## II. RELATED WORKS

Through years, the development of biometric based authentication protocol has been grown. In this section, we take a look back on some important milestones from the traditional username/password to the state-of-the-art of the remote authentication methods.

The pioneer in the field of remote authentication based password is Lamport in 1981 [4]. However, in the following works, other researchers found out the weakness of his model which made the authentication system very insecure. In Lamport's model, the users' passwords were stored directly in the server's database; therefore it was very convenience for the administrator to know or even to steal users' passwords.

In 2000, Min-Shiang Hwang and Li-Hua Li proposed a new remote user authentication scheme using smart cards [5]. The approach based smart cards has had its own advantages and disadvantages:

- The smart cards allowed business transactions to perform securely and efficiently with less interference from human. Smart cards also could be used for identification. They could provide complete identification in certain industries: car's license, patient card in eHealth, citizen card in eGovernment, student card, or bank card, … They were as easy to use as a credit card, but more secure. They were also lightweight, so easy to carry.

- The first disadvantage of smart cards was easily lost because of its small size. The smart cards could have multiple uses, therefore, the lost cards could caused a lot of troubles to their users. The second disadvantage was their level of security. They were not as secure as some people believed. This created a false sense of security, then people could not have enough care about protecting their card and its detail. In addition, there existed some possible risks of identify theft.

In 2007, Julien Bringer et al [6] introduced new approach to guarantee the security in biometric-based authentication protocol. The proposal relied on Goldwasser-Micali encryption scheme to conceal the users' biometric features; however there still existed some vulnerabilities which made system not resist the attacks from inside. Moreover, using Goldwasser-Micali was extremely time consuming, especially when the number of users increased.

In 2009, Chun-I Fan and Yi-Hui Lin [7] tried to improve the security level of biometric based remote authentication system by using multi factors authentication. The security was indeed increased, however users felt inconvenient when they were requested to provide many factors to use service. Moreover, attackers totally could impersonate users to login and to use the service on behalf of users in case they stolen all the registered factors.

In 2010, Maneesh Upmanyu, Anoop M. Namboodiri, Kannan Srinathan, and C. V. Jawahar introduced a new concept – blind authentication [3] for remote authentication protocol. This protocol was called blind for a reason that it revealed nothing but user's identity to server. The security seemed to be flawless except there was necessary to exist a Trust Third Party who took the responsibility to register and classify users. However, there was nothing to ensure that this Trust Third Party was resistant against some attacks.

In 2010, Li and Hwang [8] proposed another biometric-based remote authentication scheme using smart cards. After that in 2011, Das [9] pointed out that scheme had some flaws. And then, he proposed an enhancement of biometric based authentication protocol using smart cards in order to remedy the flaws of the previous scheme. Unfortunaly, in 2012, Cheng-Chi Lee and Chi-Wei Hsu found out that Das's scheme was vulnerable to privileged insider attacks, off-line password guessing attacks and also could not provide user anonymity. To solve these problems, Lee and Hsu presented a secure biometric-based remote user authentication with key agreement scheme using extended chaotic maps of the Chebyshev polynomials.

In 2015, Zhang Min et al pointed out some design flaws in Lee-Hsu's protocol and proposed an enhanced scheme based on fuzzy extractors in [10]. Nonetheless, these protocols did not mention the resistance to the inside attacks.

In this work, we propose a secure biometric-based remote authentication protocol using Chebyshev polynomials and Fuzzy Extractor. The fuzzy extractor scheme is applied to protect the original biometric template. The Chebyshev polynomials are used as a trapdoor which conceals a secret factor of user, however, server can be sure that this factor belongs to the user without discovering the value of this factor, and vice versa. In this scheme, we do not employ smart cards however, the security is still guaranteed in high level.

## III. PRELIMINARIES

### A. Chebyshev Polynominals

The Chebyshev polynominals and their properties, which are applied in the proposed protocol, are briefly described in this section. We call $T_n(x)$ as a Chebyshev polynomial in $x$ of degree $n$, in which $n$ is an interger. If $x$ is a variable taking value over the interval $[-1, 1]$, $T_n(x)$ will take the value over the interval $[-1, 1]$. The trigonometric definition of the Chebyshev polynomials can be presented as following:

$$T_n(x) = \cos(n \cdot arcos(x))$$

Or $$T_n(x) = \cos(n \cdot cos^{-1}(x))$$

The Chebyshev polynomials can also be defined by the recurrence relation

$$T_n(x) = 2xT_{n-1}(x) - T_{n-2}(x), \quad n \geq 2$$
$$T_0(x) = 1$$
$$T_1(x) = x$$

The Chebyshev polynomials have two important properties [11]

- Semi group property:

$$T_r(T_s(x)) = \cos\left(r \cdot cos^{-1}(\cos(s \cdot cos^{-1}(x)))\right)$$
$$= \cos(rs \cdot cos^{-1}(x)) = T_{rs}(x)$$
$$= \cos\left(s \cdot cos^{-1}(\cos(r \cdot cos^{-1}(x)))\right) = T_s(T_r(x))$$

where $r$ and $x$ are positive numbers and $x \in [-1, 1]$

This semigroup property was also defined on interval $(-\infty, +\infty)$ by Zhang [12] in 2008. Therefore, the property can be represented as follows:

$$T_n(x) \equiv \left(2xT_{n-1}(x) - T_{n-2}(x)\right) \bmod p, \quad n \geq 2$$

Where $p$ is a large prime. Evidently,

$$T_r(T_s(x)) \equiv T_{rs}(x) \equiv T_s(T_r(x)) \bmod p$$

This is effective property, very useful in authentication protocol, then we apply it in our proposal as public key cryptosystem.

- The chaotic property:

When the degree $n > 1$, the Chebyshev polynomial $T_n(x): [-1, 1] \to [-1, 1]$ is a chaotic map with its invariant density $f^*(x) = 1/\pi\sqrt{1-x^2}$, for positive Lyapunov exponent $\lambda = \ln(n)$.

The security level of the Chebyshev polynomials depend on two typical problems, which are assumed to be difficult to solve in polynomial time:

(1) The Discrete logarithm problem (DLP): Given two elements $a$ and $b$, there is a computational problem of finding $x = \log_a b$ such that $b = a^x$. In Chebyshev speaking, it is hard to find the integer $x$, such that $T_x(a) = b$.

(2) The Diffie-Hellman problem (DHP): Given an element $x$, and the value of $g^x$ and $g^y$, how to compute the value of $g^{xy}$, in which $x, y$ are two randomly chosen integer, and $g$ is a generator of some group (it can be a function). Therefore, in Chebyshev speaking, though we know three element $x$, $T_r(x)$ and $T_s(x)$, it is difficult to calculate the value of $T_{rs}(x)$

### B. Secure Sketch and Fuzzy Extractor

Secure sketch is a technique which allows a noisy input to be reconstructed. It has two components, SS – Sketch and Rec – Recover. SS component takes an input w, then returns a sketch s. To Rec component, if its input w' is close to w and there is the helper data s, it can return exactly w. The key point of the secure sketch is the public sketch s does not disclose the biometric information w.

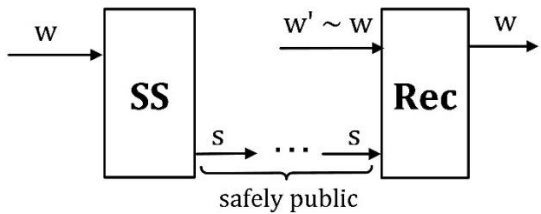

Fig. 1. The secure sketch scheme.

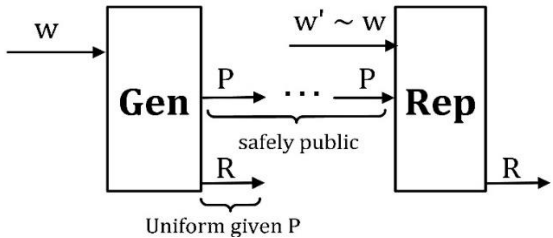

Fig. 2. The fuzzy extractor scheme

As shown in Fig. 2, fuzzy extractor is a technique which can extract nearly uniform randomness R from its input. This extraction is error-tolerant since R is unchanged even if the input is not the same, as long as the gap between it and the previous one is acceptable. Consequently, R is used as a key to encrypt/decrypt or to authenticate. In other words, fuzzy extractor is counted as biometric tool to authenticate a user using his or her own biometric template as a key.

To preserve high security, the authors in [13] applied secure sketch to construct fuzzy extractor. The scheme is demonstrated as Fig. 3.

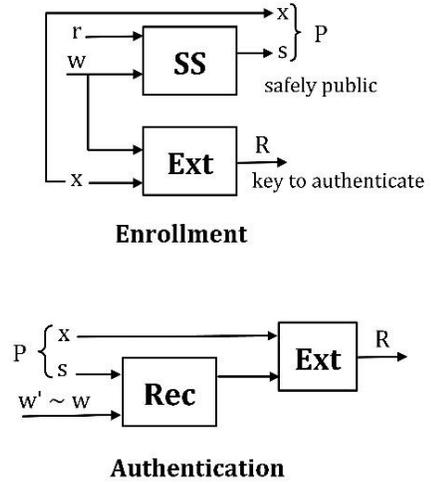

Fig. 3. Construction of Fuzzy Extractors from Secure Sketches [13]

## IV. PROPOSAL PROTOCOL

TABLE I. NOTATIONS USED IN PROPOSED PROTOCOL

| Notation | Descripstion |
| --- | --- |
| K | Key generated from Fuzzy Extractor process |
| HD | Helper Data generated from Secure Sketch scheme |
| $PW_T$ | User's password created in the enrollment phase |
| $B_T$ | The original biometric template |
| $BB_T$ | The secure version of biometric template |
| PW | The password provided in authentication phase |
| B | The biometric data provide in authentication phase |
| K | The biometric key generated in authentication phase |
| p | A large prime number |
| ID | User identity |
| $R_C, R_S$ | The random integers generated from Client and Server |
| h(.) | A secure one way hash function |
| $A \oplus B$ | The exclusive-or (XOR) operation between A and B |
| A // B | The concatenation operation between A and B |
| t | The time-stamp |

## A. Enrollment Phase

In the enrollment phase of this client/server architecture, a client side definitely makes a first move. At first, a user chooses his/her password $PW_T$, then provides his/her biometric feature $B_T$ (for example: user's face) for a sensor. The Fuzzy Extractor module installed in the user's device takes $PW_T$ and $B_T$ as inputs to generate a biometric key $K_T$ in the first step. Next step, user calculates the secure biometric version $BB_T = B_T \oplus K_T$ and $h(K_T \| PW_T)$. Then, he/she sends those two elements to server side in the third step. Note that all messages between the client and the server over transmission network are protected by asymmetric cryptosystem (PKI – Public Key Infrastructure)

Once receiving the package from user, server selects randomly two number $s$, $X_S$ ($X_S$ is kept in secret), then calculates $SPUB \equiv T_{X_S}(s) \bmod p$. $SPUB$ is considered as a public key of registration server. In the next step, server creates user's identity $ID$ and stores all these factors (including $ID$, $BB_T$, $X_S$, $s$) into the database. After that, server continues to calculate $O_1$, and $O_2$ as follows:

$$O_1 = h(BB_T \| ID)$$
$$O_2 = h(K_T \| PW_T) \oplus h(X_S)$$

Next, server sends to user a package including $O_1$, $O_2$, $s$, $SPUB$, $p$. Once receiving, user stores all these factors into his/her device. The enrollment phase is illustrated as Fig. 4

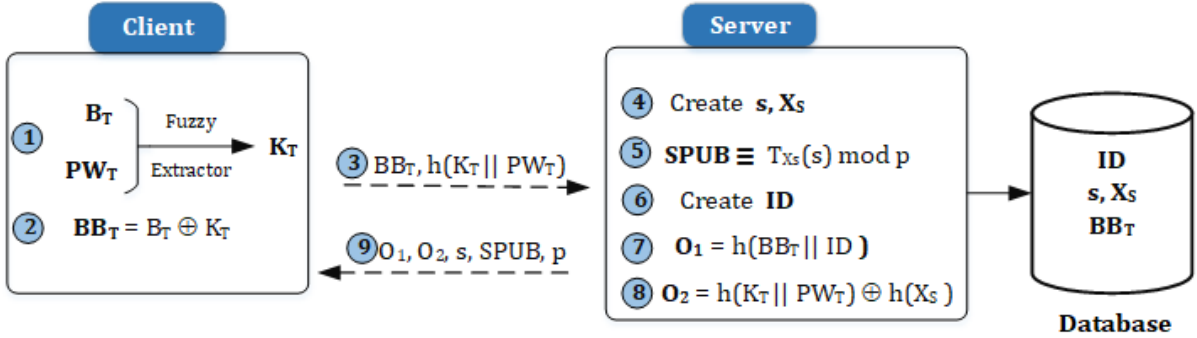

Fig. 4. Enrollment Phase

## B. Authentication Phase

In this phase, user needs to provide his/her biometric feature B and retype the registered password PW. The Fuzzy Extractor module in his/her device generates the authentication key K from these inputs. In the next step, the newly generated key K combines with B to create the secure biometric version BB of the authentication phase.

$$BB = B \oplus K$$

In the mean of time, user create a random number $R_C$, then calculates the following messages:

$$M_1 = T_{R_C}(s)$$
$$M_2 = T_{R_C}(SPUB) = T_{R_C}(T_{X_S}(s))$$
$$\alpha = h(K \| PW) \oplus h(M_1 \| M_2 \| t_1)$$

After that, user sends to server the package including $O_1, O_2, BB, M_1, \alpha, t_1$ in the fourth step.

At server side, server has to perform some steps to verify its client. First of all, server checks timestamp $t_1$. If the difference between $t_1$ and current time is less than the predefined threshold, server will do the next steps; if not, the process will be terminated. Server calculates $O'_1 = h(BB \| ID)$ from the received $BB$ and the $ID$ stored in the database, then compares the result with the received $O_1$. If matched, it also proves that this user has registered to the server and his/her biometric feature are matched. Server continues to do some calculations:

$$M'_2 = T_{X_S}(M_1) = T_{X_S}(T_{R_C}(s))$$

$$temp = O_2 \oplus h(X_S) = h(K_T \| PW_T) \oplus h(X_S) \oplus h(X_S)$$
$$= h(K_T \| PW_T)$$

The results are used to calculate $\alpha'$.

$$\alpha' = temp \oplus h(M_1 \| M'_2 \| t_1)$$
$$= h(K_T \| PW_T) \oplus h(M_1 \| M'_2 \| t_1)$$

Server performs the comparison between $\alpha$ and $\alpha'$. If equal, it confirms that the password and the biometric data user provided are correct. And, $M'_2$ is equal to $M_2$, it helps to verify the $SPUB$ without revealing $R_C$ to server or $X_S$ to user.

In the next step, server randomly creates $R_S$, and do some calculations:

$$M_3 = T_{R_S}(s)$$
$$\beta = h(M'_2 \| M_3 \| t_2)$$

Then, server sends back to user the package including $M_3, \beta, t_2$. This is the first stage of the authentication phase. The purpose of this stage is to authenticate user. The following steps belongs to the second stage. The purpose of this stage is to help user authenticate the server he/she communicates with.

Once receiving the package in the eighth step, user check the timestamp $t_2$. If the difference between $t_2$ and current time is less than the predefined threshold, user will do the next steps; if not, the process will be terminated. In case $t_2$ is validated, user calculates $\beta' = h(M_2 \| M_3 \| t_2)$; then, compares $\beta$ and $\beta'$. If they are equal, it also means $M_2$ and $M'_2$ are equal, so user can be sure that the server he/she communicates with is the server

he/she registered. After that, user needs to perform some calculations:

$$M_4 = T_{Rc}(M_3) = T_{Rc}(T_{Rs}(s))$$

$$\gamma = h(M_2 \| M_4 \| t_3)$$

User sends $\gamma, t_3$ to server through the tenth step. Once receiving, server checks timestamp $t_3$ first. If $t_3$ is validated, server will do the following calculations:

$$M'_4 = T_{Rs}(M_1) = T_{Rs}(T_{Rc}(s))$$

$$\gamma' = h(M'_2 \| M'_4 \| t_3)$$

Then, server compares $\gamma$ and $\gamma'$. The fact that they are equal proves that $M_4$ and $M'_4$ are equal. Server and user can use $M_4$ as a session key for later communication with each other by using a symmetric cryptosystem. And at last, the authentication is successful and server/client achieves a session key agreement. The authentication phase is illustrated as Fig. 5.

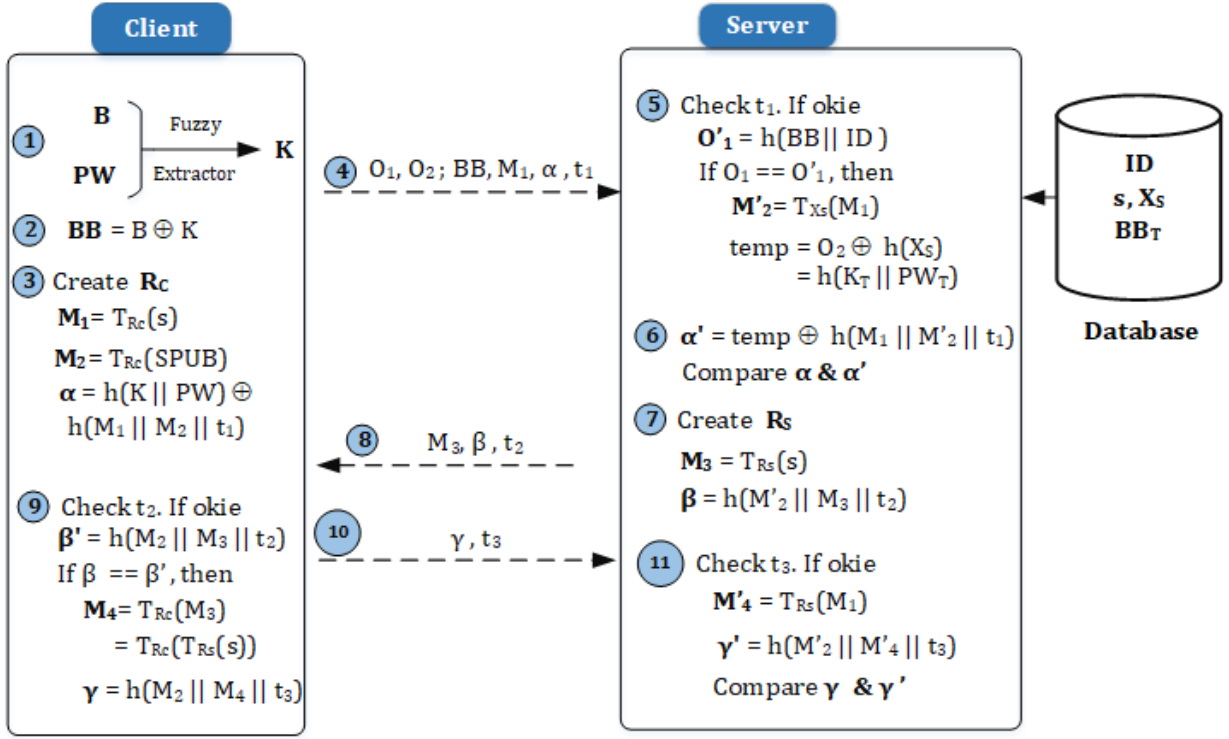

Fig. 5. Authentication phase

## V. EVALUATION

### A. Security analysis

Mutual authentication: Our protocol can achieve mutual authentication between client and server. In the authentication phase, server has to verify $\alpha$ and $\gamma$ in order to ensure this user is the one he/she declares. Moreover, user also has to verify $\beta$ to authentication server. If attackers want to forge the messages, they have to face the typical problem DLP and DHP.

Biometric template attack: the original biometric data of user is protected by the fuzzy extractor scheme. The process of generating biometric key is done in client side. The authentication factor server is known is the secure version of user's biometric data $BB_T$. Therefore, the original biometric data cannot be compromised.

Replay attack. The attacker can not reuse the old messages to deceive neither server nor client, because all these messages between client and server are added timestamps. Therefore, the attacke cannot pass the verification without the correct timestamp.

Man-in-the-middle attack. In this attack, the attacker interfere in the communication between client and server while they still believe that they are talking to each other over a private connection. In this protocol, the mutual authentication requirement is satisfied, so it is very difficult for anyone to break into communication between client and server.

User anonymity: Somehow the attacker can eavesdrop on the conversation between client and server, then, try to track the user's real identity to discover some sensity information of the user. In this protocol, the user's identity ID is protected by $O_1 =$

$h(BB_T \parallel ID)$. Therefore, our proposal can provide the user anonymity.

*B. Accuracy*

The training data contains a set of 50 face images, which are all in JPG format, 180 x 200 pixels. This set includes 22 images of 22 European and 28 images of 15 Asian.

The registration data set contains 65 face images, including: 54 images of 54 European and 11 images of 7 Asian. We divide the testing data set into 4 groups

Group A includes 65 other images of the same 61 people in the registration set. The result from this group presents the fasle reject rate (FRR)

Group B, C, D: each group includes 65 images of different people who did not register. The result from these groups present the false accept rate (FAR).

TABLE II. EXPERIMENTAL RESULT OF AUTHENTICATION PROCESS

| Group | FAR | FRR |
| --- | --- | --- |
| A | X | 5/65 |
| B | 3/65 | X |
| C | 3/65 | X |
| D | 0/65 | X |

The experimental results in Table II show that the average FAR is 3.07% and the average FRR is 7.96%. The fact that the error rates are quite low proves that the ability of face recognition of this proposal is acceptable. In the context of high level of security, this proposal absolutely can put into practice.

## VI. CONCLUSION

In this paper, we have presented a noble biometric-based remote authentication protocol to most of sophisticated attacks over an open network. The proposed protocol combines user biometric with the other authentication factors to achieve the high level of security. The Chebyshev polynomials and fuzzy extractor scheme have been employed to ensure the protocol is resistant to biometric template attack, replay attack, man-in-the-middle attack, user anonymity, and also achives the mutual authentication and session key agreement at the end of the authentication phase. The fact that the accuracy of this facial regconition system is quite good makes us believe that it can be put into practice in the near future.

ACKNOWLEDGMENT

This research is funded by Vietnam National University - Ho Chi Minh City (VNU-HCM) under grant number T-PTN-2017-88. We also want to show a great appreciation to each member of D-STAR Lab (www.dstar.edu.vn) for their enthusiastic supports and helpful advices during the time we have carried out this research.